\documentclass[12pt]{article}
\usepackage{graphicx,psfrag,amsmath,calc,amssymb}
\begin{document}

\pagestyle{headings}
\title{\bf{Coexistence of Antiferromagnetism and Triplet Superconductivity}}
\author{Wei Zhang and C. A. R. S\'{a} de Melo\\
\\
{\it School of Physics, Georgia Institute of Technology,
             Atlanta, Georgia 30332}}
\maketitle
\newpage

%

The authors discuss the possibility of coexistence of antiferromagnetism and triplet superconductivity 
as a particular example of a broad class 
of systems where the interplay of magnetism and superconductivity 
is important. This paper focuses on the case of quasi-one-dimensional metals,
where it is known experimentally that antiferromagnetism
is in close proximity to triplet superconductivity in the temperature
versus pressure phase diagram. Over a narrow range of pressures, the authors 
propose an intermediate non-uniform phase consisting of alternating
insulating antiferromagnetic and triplet superconducting stripes.
PACS numbers: 74.70.Kn, 74.25.DW
\newpage

(TMTSF)$_2$PF$_6$ is a quasi-one dimensional organic material with a 
complex phase diagram involving antiferromagnetic (AF) insulating and 
triplet superconducting (TS) phases~\cite{lee-00, lee-02a, lee-02b}. 
New experiments on 
this material in high magnetic fields have
shown that TS phase is strongly affected
by the proximity to AF phase characterized by a 
spin density wave (SDW)~\cite{lee-02a}. 
Motivated by these
experiments and the known phase diagram of
quasi-one-dimensional ${\rm (TMTSF)_2 PF_6}$ we propose a new 
phase for quasi-one-dimensional systems where AF (SDW)
and TS coexist. 
The coexistence of these phases implies that the new state 
is non-uniform, with alternating stripes of insulating 
AF and TS, due to the appearance of a negative
interface energy between AF and TS regions. 
As indicated in the schematic 
phase diagram (Fig. 1),
the inhomogeneous intermediate phase is expected to
exist over a narrow range of pressures
$\Delta P = P_2 (T)  - P_1 (T)$ around $P_c$, where $\Delta P \ll P_c$. 
\begin{figure}
\begin{center}
\psfrag{P1(T)}{\scriptsize{$P_1(T)$}}
\psfrag{P2(T)}{\scriptsize{$P_2(T)$}}
\psfrag{P1}{\scriptsize{$P_1$}}
\psfrag{P2}{\scriptsize{$P_2$}}
\psfrag{P3}{\scriptsize{$P_3$}}
\psfrag{Pc}{\scriptsize{$P_c$}}
\psfrag{Tc}{\footnotesize$T_c$}
\psfrag{P1(0)}{\scriptsize$P_1(0)$}
\psfrag{P2(0)}{\scriptsize$P_2(0)$}
\psfrag{(TMTSF)2PF6}{\scriptsize$\textrm{(TMTSF)}_2\textrm{PF}_6$}
\includegraphics[width=6.75cm]{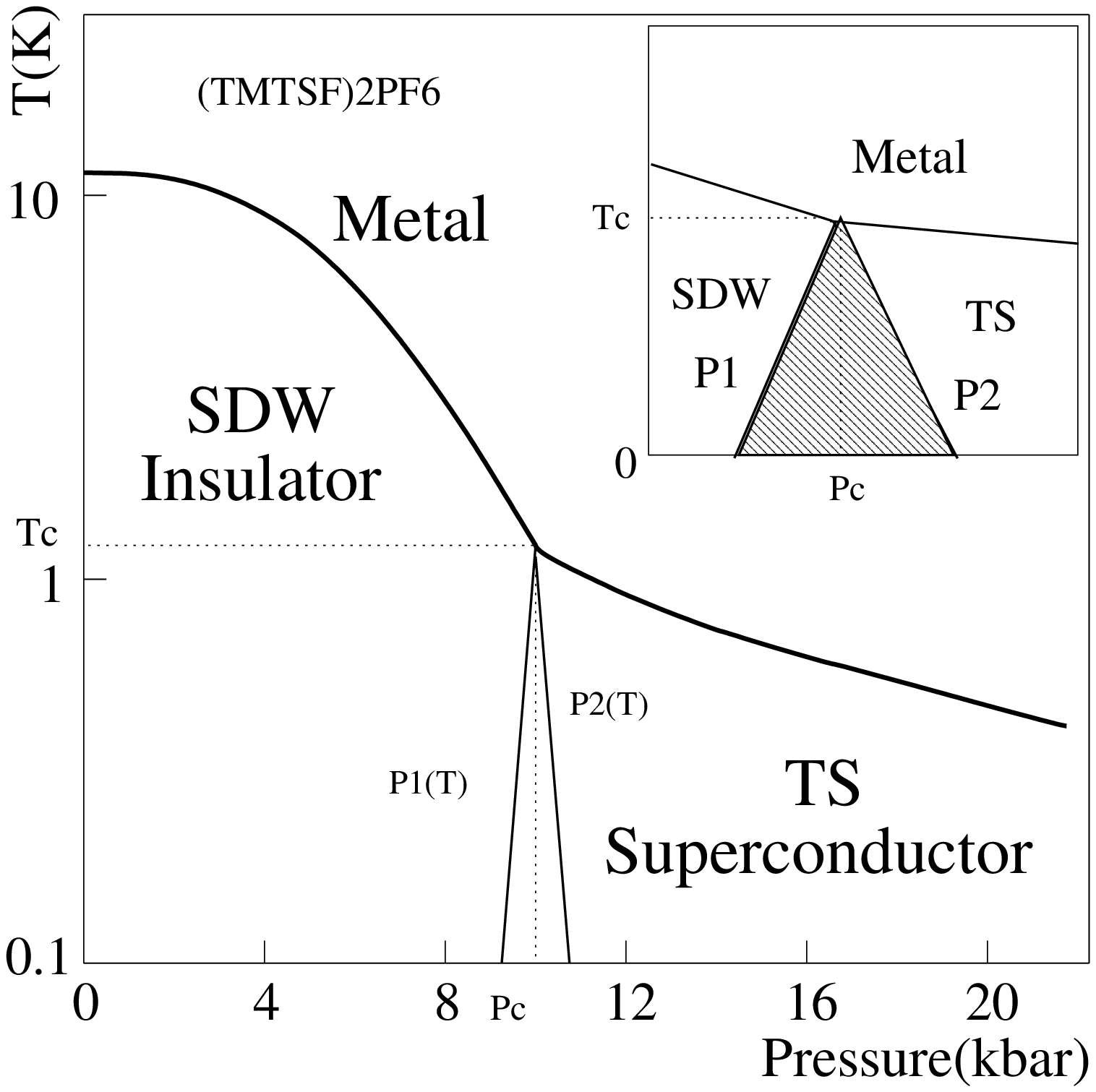}
\psfrag{lAF}{$\ell_{AF}$}
\psfrag{lTS}{$\ell_{TS}$}
\psfrag{Sb}{{\footnotesize$S_b$}}
\psfrag{Dc}{\footnotesize$D_c$}
\psfrag{L}{\footnotesize$L$}
\psfrag{Superconductor}{\footnotesize Superconductor}
\psfrag{Insulator}{\footnotesize Insulator}
\includegraphics[width=6.75cm]{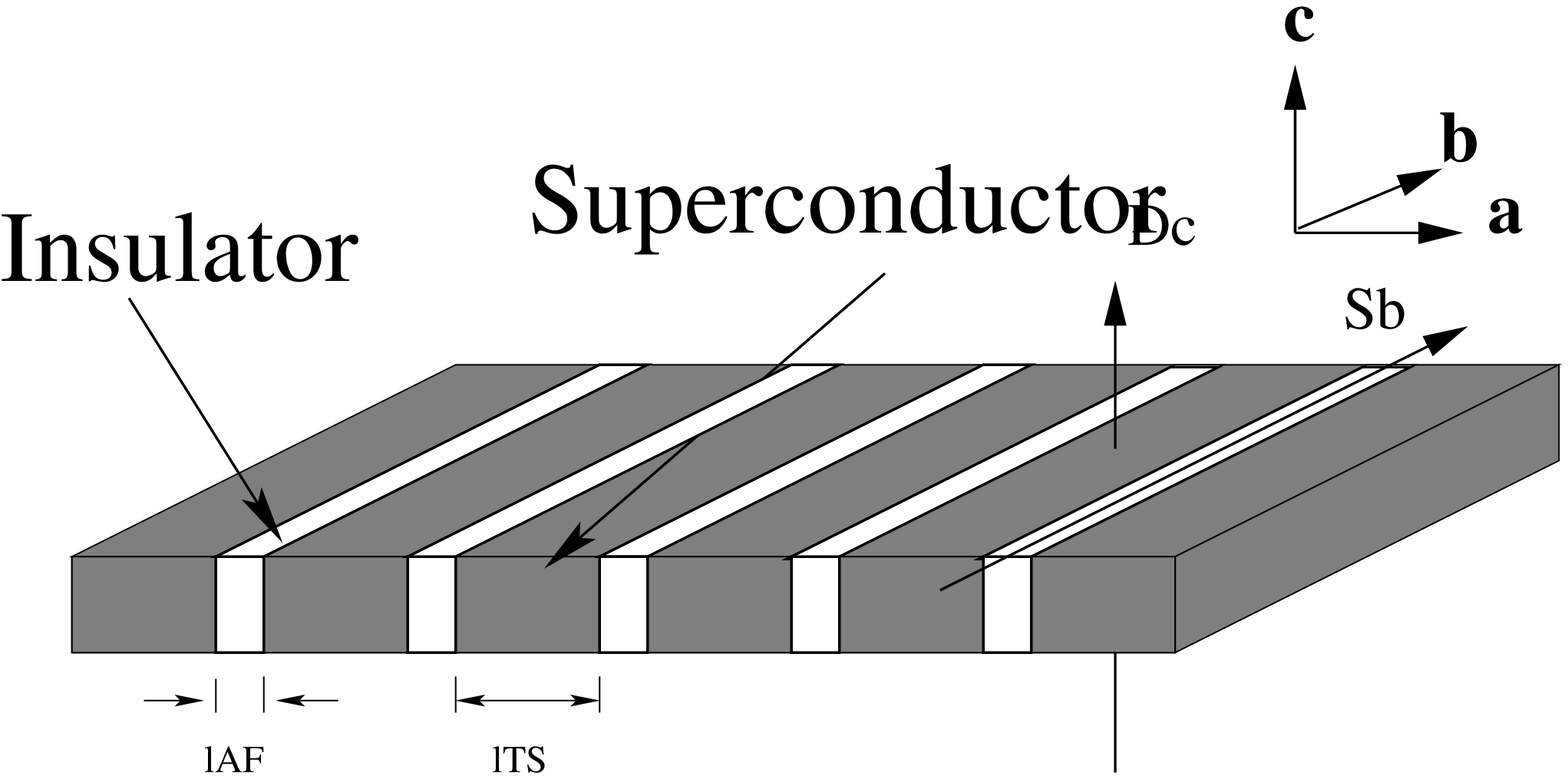}
\end{center}
\caption{ a) Phase diagram of ${\rm (TMTSF)_2PF_6}$
in a log-linear scale (from~\cite{jerome-82}), showing schematically the  
proposed AF-TS coexistence region (inset).
b) Schematic drawing of the proposed stripe pattern for the AF-TS 
coexistence region.}
\label{fig:phase-diagram}
\end{figure}

{\it Effective Free Energy:}
The possibility of coexistence of SDW and TS in quasi-one-dimensional conductors
transcends microscopic descriptions based on 
standard g-ology, where SDW and TS phase boundaries 
neighbor each other but do not coexist~\cite{solyom-79}. 
Inspired by experiments~\cite{lee-02b,mortensen-82}, 
we model ${\rm (TMTSF)_2 PF_6}$ as a highly anisotropic orthorombic crystal, 
and we take the primary 
directions of the SDW vector order parameter to be the b-axis (y-direction),
and the primary direction of the TS vector order parameter to
be the c-axis (z-direction).
Furthermore, we consider the spatial variation of the 
SDW or TS order parameter to be along the a-axis (x-direction),
as a reflection of the quasi-one-dimensionality of the system.
This simplifies the choices of the order parameters 
to be ${\bf S} ({\bf r}) \to S_b (x)$, and 
${\bf D} ({\bf r}) \to D_c (x)$, and reduces the associated 
effective field theory to one spatial dimension.
 
Thus, the generalized Ginzburg-Landau free energy in real space can 
be written as
\begin{equation}
\label{eqn:f-tot}
\mathcal{F}_{tot}  = 
\mathcal{F}_{AF}  + 
\mathcal{F}_{TS}  + 
\mathcal{F}_{C},   
\end{equation}
where $\mathcal{F}_{AF}$,  $\mathcal{F}_{TS}$, 
$\mathcal{F}_{C}$ are the AF(SDW) , TS and coupling
contributions discussed below.   
The AF contribution is
\begin{equation}
\label{eqn:AF-FreeEnergy}
\mathcal{F}_{AF}  = 
\int_{L_{AF}}dx 
\left[ 
U_{AF} (x) + V_{AF} (x)
\right],
\end{equation}
where 
$
U_{AF}(x) = {\alpha_{AF}}{\lvert S_b(x)\rvert}^2 +
{\beta_{AF}}{ \lvert \partial_x S_b(x) \rvert }^2 +
{\gamma_{AF}}{\lvert\ S_b(x)\rvert}^4 
$
represents a typical GL free energy density, and
$
V_{AF} = 
{\delta_{AF}}{\lvert \partial_x S_b(x) \rvert}^4 +
{\theta_{AF}}{\lvert S_b(x) \rvert}^2 {\lvert \partial_x S_b(x) \rvert}^2     
$
represents the extra terms in the expansion, which are
relevant close to $P_1 (T)$. 
The TS contribution is
\begin{equation}
\label{eqn:TS-FreeEnergy}
\mathcal{F}_{TS}  = 
\int_{L_{TS}}dx 
\left[ 
U_{TS} (x) + V_{TS} (x)
\right],
\end{equation}
where 
$
U_{TS}(x) = {\alpha_{TS}}{\lvert D_c(x)\rvert}^2 +
{\beta_{TS}}{ \lvert \partial_x D_c(x) \rvert }^2 +
{\gamma_{TS}}{\lvert\ D_c(x)\rvert}^4 
$
, and
$
V_{TS} = 
{\delta_{TS}}{\lvert \partial_x D_c(x)\rvert}^4 +
{\theta_{TS}}{\lvert D_c(x) \rvert}^2 {\lvert \partial_x D_c(x) \rvert}^2     
$
. To describe the coexistence region
the two order parameters must couple. 
To conform with independent Parity invariance, 
\begin{equation}
\label{eqn:Free-Energy-Coupling-Local}
\mathcal{F}_{C} = 
\sum_{inter} 
\int_0^{\ell_p} dx \lambda_{bc}'
\lvert S_{b} (x)   \rvert^2
\lvert D_{c} (x)   \rvert^2,
\end{equation}
where the sum is over all possible interfaces between AF and
TS, the coupling constant $\lambda_{bc}'$ is pressure
and temperature dependent, and $\ell_p$ is the length of proximity at which 
AF and TS order parameters coexist locally. 
Since the AF order is pair breaking 
to triplet electron pairs \cite{ketterson-song-98, nakajima-73}, 
we estimate that $\ell_p$ to be much smaller 
than the length of the AF stripe $\ell_{AF}$, 
unless $P \to P_2(T)$, 
where $\ell_{AF}$ approaching zero. 
In this limiting case, 
the Josephson effect between two consecutive TS stripes becomes significant and 
the material will change from a 2D to 
a 3D superconductor. 
In the case where $\ell_p$ is small, 
both AF and TS order parameters inside the proximity region can be 
approximated by linear functions, thus 

\begin{equation}
\label{eqn:Free-Energy-Coupling}
\mathcal{F}_{C} = 
\sum_{inter}
\lambda_{bc} 
\lvert \partial_x  S_{b} (x)   \rvert^2
\lvert \partial_x  D_{c} (x)   \rvert^2,
\end{equation}
where 
$\lambda_{bc} = \lambda_{bc}' \int_0^{\ell_p} dx x^2 (\ell_p -x)^2$ 
is the new coupling constant. 

{\it Saddle Point Equations:}
To obtain the saddle point equations, we minimize $\mathcal{F}_{tot}$
with respect to $S_b (x)$ and $D_c^* (x)$. 
Variation of $\mathcal{F}_{tot}$ with respect 
to  $S_b (x)$ 
lead to the differential equation
$$
[
2\alpha_{AF} 
+ 
4 \gamma_{AF} S_b^2(x)
- 
\beta_{AF} \partial^2_x
] S_b(x) 
+
{\hat M}_{AF} S_b(x) 
= 0, 
$$
with 
$
{\hat M}_{AF} S_b(x) =
-
\delta_{AF} \partial_x \left[(\partial_x S_b(x)\right]^3 
+ 
2 \theta_{AF} S_b(x) \lvert \partial_x S_b(x) \rvert^2.
$
Variation of $\mathcal{F}_{tot}$ with
respect to $D_c^* (x)$ lead to a similar equation.
Given that we are considering the possibility of coexistence of 
the two phases, the boundary conditions in the presence of 
AF-TS interfaces require that $S_b (x)\vert_{inter^+} = 0$ and 
$D_c (x)\vert_{inter^-} = 0$, where $inter^+$ and $inter^-$ denote the 
two boundaries of locally coexisting proximity $\ell_{p}$ at which 
$S_b(x)$ and $D_c(x)$ vanish respectively.

{\it Variational Free Energy:} 
We consider first the AF case and search for periodic solutions with
period $\ell_{AF}$, 
with $S_b (x)\vert_{inter^+} = 0$ at 
the AF-TS interfaces. For a given volume of the AF region, controlled
by $L_{AF}$, the Free energy associated with the AF phase becomes
the sum of $N_{AF}$ identical terms, where $N_{AF} = L_{AF}/\ell_{AF}$ 
gives the number of AF stripes. 
Generally, each term in $\mathcal{F}_{AF}$ corresponds to an 
insulating AF stripe characterized by the order parameter 
$S_b (x) = \sum_n A_n \sin (Q_n x)$, where $Q_n = 2\pi n/\ell_{AF}$.
But here we take for simplicity the variational class where
$S_b (x) = A_1 \sin (Q_1 x)$, which satisfies the appropriate boundary
and merging conditions. To simplify notation, we use
$A_1 \to A$ and $Q_1 \to Q$.
In this case
\begin{equation}
\label{eqn:free-af-variational}
\mathcal{F}_{AF}
=
{L_{AF}} \Big[
C_2(Q) A^2 + C_4(Q) A^4
\Big],
\end{equation}
where
$
C_2 (Q) = 
(\alpha_{AF} + \beta_{AF}Q^2)/2
$
and  
$
C_4(Q) = 
( 3 \gamma_{AF} + \theta_{AF} Q^2 + 3 \delta_{AF} Q^4 )/8.
$

The same type of analysis applies to $\mathcal{F}_{TS}$.
In the absence of a magnetic field,
we assume periodic solutions of the form 
$D_c (x) = B \sin (K x)$,
Here $B$ can be complex, but independent
of position $x$, making this choice consistent with 
a weak spin-orbit coupling interaction from a microscopic
theory~\cite{duncan-01}. All the analysis discussed for the AF case
applies with the following change of notation:
$L_{AF} \to L_{TS}$, $A \to B$, 
$Q \to K$, $\alpha_{AF} \to \alpha_{TS}$,
$\beta_{AF} \to \beta_{TS}$, etc. 
And the coupling free energy is
\begin{equation}
\label{eqn:free-inter-variational}
\mathcal{F}_C 
=
N_{int} \Lambda (Q, K) A^2 |B|^2,
\end{equation}
where 
$N_{int} = 2 N$ is the total number of interfaces, 
$ \Lambda (Q, K) A^2 |B|^2 = f_{int}$ is the
free energy of one interface with  
$\Lambda (Q, K) = \lambda_{bc} Q^2 K^2$. 

{\it Variational Solution:} 
Variations of $\mathcal{F}_{tot}$ with respect to $\phi_{AF}=A$, 
$\phi_{TS}=|B|$,  and $q_{AF}=Q$ or $q_{TS}=K$
lead to the non-trivial solutions
\begin{equation}
\label{eqn:phi-i}
 \phi_i^2  = 
 \frac
{ 4 \beta_i \theta_i  - 24 \alpha_i \delta_i }
{ 36 \gamma_i \delta_i - \theta_i^2 },
\end{equation}
\begin{equation}
\label{eqn:q-i}
 q_i^2 =
\frac 
{ \alpha_i \theta_i - 6 \beta_i \gamma_i }
{ \beta_i \theta_i - 6 \alpha_i \delta_i }.
\end{equation}
In addition, $f_{int} = 
\lambda_{bc} \phi_{AF}^2 \phi_{TS}^2 q_{AF}^2 q_{TS}^2$.  and 
the width of each stripe is given by
\begin{equation}
\label{eqn:l-i}
\ell_{i} = 2 \pi 
\sqrt{
\frac {\beta_i \theta_i - 6 \alpha_i \delta_i}
{ {\alpha_i} \theta_{i} - 6 \beta_{i} \gamma_{i} } 
},
\end{equation}
where $i = AF, TS$. If $\lambda_{bc} (P,T) > 0$, 
the system phase separates, and there is no coexistence region,
thus the line separating the AF phase from the TS phase indicates
a discontinuous transition and $(P_c, T_c)$ is bicritical.
However, if $\lambda_{bc} (P,T)  < 0$ then the formation of interfaces
is preferred, and alternating stripes of AF and TS order
appear in the system, creating a coexistence region
dictated by the condition $\lambda_{bc} (P,T) < 0$.
In this case, two additional
transition lines emanate from $(P_c, T_c)$.
This indicates that the point $(P_c, T_c)$
in the phase diagram illustrated in Fig.~1 can be bicritical, tricritical,
or tetracritical and corresponds to the place where $\lambda_{bc} (P,T) = 0$.
\begin{figure}
\begin{center}
\psfrag{Pressure}{Pressure}
\psfrag{P1}{\scriptsize$P_1$}
\psfrag{P2}{\scriptsize$P_2$}
\psfrag{Pc}{\scriptsize$P_c$}
\psfrag{Energy (N(EF)Tc2)}{Energy ({\scriptsize$N(\epsilon_F)T_c^2$})}
\includegraphics[width=6.4cm]{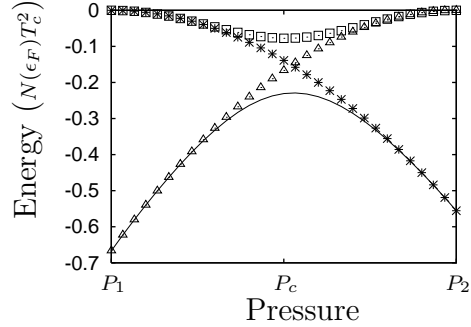}
\end{center}
\caption{
Free energies for the coexistence and the
pure AF and TS phases. Solid line $\to$
$\mathcal{F}_{tot}$; stars $\to$ $L f_{TS}$; 
triangles $\to$ $L f_{AF}$;
squares $\to$ $\mathcal{F}_{C}$. 
Exponents chosen are 
$\varepsilon_{\alpha_i} = 5.0$, 
$\varepsilon_{\beta_i} = 1.0$, $\varepsilon_{\theta_i} = 5.0$, $\varepsilon_{\lambda_i} = 1.0$
where $i = {AF,~TS}$ and 
dimensionless parameters: 
$\tilde\alpha_1 = \tilde\beta_1  = 1.0$,
$\tilde\gamma_1 = 0.05$,
$\tilde\delta_1 = 0.005$,
$\tilde\theta_1 = 0.001$;
and 
$\tilde\alpha_2 = \tilde\beta_2  = 1.0$,
$\tilde\gamma_2 = 0.07$,
$\tilde\delta_2 = 0.006$,
$\tilde\theta_2 = 0.002$, 
$\tilde\lambda = 0.004$. 
}
\label{fig:length-order-parameter}
\end{figure}

{\it Phase Transitions:} 
Starting from the point $(P_c, T_c)$, 
two transition lines appear. The transition 
line $P_1 (T)$ corresponds to the disappearance of the pure AF 
phase, and the transition line $P_2 (T)$ corresponds to the appearance
of the pure TS phase. (We estimate the Josephson effect region 
is small and look $P_2(T)$ and $P_3(T)$ as one single line.) 
For pressures between $P_1 (T)$ and $P_2 (T)$
there is coexistence between AF and TS order in the form of stripes.
This implies that at $P_1 (T)$ the TS stripe width $\ell_{TS} = 0$,
while at $P_2 (T)$ the AF stripe width $\ell_{AF} = 0$. 
Furthermore, for 
$P_1 (T) < P < P_2 (T)$, $\ell_{TS}$ increases
from 0 to some finite value and $\ell_{AF}$ decreases from some other finite value 
to 0 with increasing pressure. In order to meet these 
and the saddle point requirements, the 
parameters appearing in $\mathcal{F}_{tot}$ must behave as follows. 
We define the reduced pressure changes $\Delta P_i = [P - P_i (T)]/P_c$, 
where $i =1,2$ to analyse the AF and TS
parameters. For $P <  P_2 (T)$, the AF parameters have the form
$\gamma_{AF} = \gamma_1 N(\epsilon_F)T_c^2 > 0$; 
$\delta_{AF} = \delta_1  N(\epsilon_F)T_c^2 > 0$;
$\alpha_{AF} = \alpha_1 N(\epsilon_F)T_c^2 
\vert \Delta P_2 \vert^{\varepsilon_{\alpha_{AF}}}$,
with $\alpha_1 < 0$;
$\beta_{AF} = \beta_1 N(\epsilon_F)T_c^2 \vert \Delta P_2 \vert^{\varepsilon_{\beta_{AF}}}$,
with $\beta_1 < 0$;
$\theta_{AF} = \theta_1 N(\epsilon_F)T_c^2 
\vert \Delta P_2 \vert^{\varepsilon_{\theta_{AF}}}$,
with $\theta_1 < 0$; and $36\gamma_{AF} \delta_{AF} - \theta_{AF}^2 > 0$.
For $P > P_1 (T)$,
the TS  parameters have the form
$\gamma_{TS} = \gamma_2 N(\epsilon_F)T_c^2 > 0$; 
$\delta_{TS} = \delta_2 N(\epsilon_F)T_c^2 > 0$;
$\alpha_{TS} = \alpha_2 N(\epsilon_F)T_c^2 \vert \Delta P_1\vert^{\varepsilon_{\alpha_{TS}}}$,
with $\alpha_2 < 0$;
$\beta_{TS} = \beta_2 N(\epsilon_F)T_c^2 \vert \Delta P_1 \vert^{\varepsilon_{\beta_{TS}}}$,
with $\beta_1 < 0$;
$\theta_{TS} = \theta_2 N(\epsilon_F)T_c^2 
\vert \Delta P_1 \vert^{\varepsilon_{\theta_{TS}}}$,
with $\theta_2 < 0$; and $36\gamma_{TS} \delta_{TS} - \theta_{TS}^2 > 0$.
Consider now, the interface terms in the region
$P_1 (T) < P < P_2 (T)$, which has the form
$\lambda_{bc} = \lambda_0 N(\epsilon_F)T_c^2
{\rm sgn} [(P - P_1) (P - P_2)] 
\vert \Delta P_1 \vert^{\varepsilon_{\lambda_{AF}}}
\vert \Delta P_2 \vert^{\varepsilon_{\lambda_{TS}}},
$
with $\lambda_0 > 0$.
This form is required to make the interface
energy negative between $P_1 (T)$ and $P_2 (T)$.

Next, we focus only on the analysis of $\ell_{TS}$ 
and $\mathcal{F}_{tot}$ in the vicinity of
$P_1(T)$. 
The requirement that $\ell_{TS} \to 0$ 
as $P \to P_1(T)$ forces a constraint 
$\varepsilon_{\alpha_{TS}} >  \varepsilon_{\beta_{TS}}$. 
By considering $\mathcal{F}_{tot}$ must less than 
the pure phase $\mathcal{F}_{AF}(P_1)$, 
another condition $\varepsilon_{\alpha_{TS}} \ge 
3 \varepsilon_{\beta_{TS}} + 2 \varepsilon_{\lambda_{TS}}$ 
is imposed. 
Furthermore, calculation of $\partial \mathcal{F}_{tot}/ \partial P$ 
shows the phase transition at $P_1$ is continuous if 
$\varepsilon_{\lambda_{TS}}+ \varepsilon_{\beta_{TS}} > 1$ or 
discontinuous if $\varepsilon_{\lambda_{TS}}+ \varepsilon_{\beta_{TS}} \le 1$. 
 In Fig.~2, we show the behavior of the various contributions to
$\mathcal{F}_{tot}$ for the case where the transitions are
continuous at $P_1$ and $P_2$, and thus $(P_c, T_c)$ is tetracritical.
Dimensionless parameters used are defined as
$\tilde\alpha_i = \tilde\beta_i =  \rho_i^{1/2}$,  
$\tilde\gamma_i = \gamma_i \sigma_i^{-1/2}$,
$\tilde\delta_i = \delta_i \sigma_i^{3/2}$,
$\tilde\theta_i = \theta_i \sigma_i^{1/2}$,
and 
$\tilde\lambda_0 = \lambda_0 \sigma_1 \sigma_2$,
where $\rho_i = \alpha_i \beta_i$,
and $\sigma_i =  \alpha_i/\beta_i$, with $i = 1, 2$. 
We note in passing that the analysis of $\ell_{AF}$ and $\mathcal{F}_{tot}$ 
in the vicinity of $P_2(T)$ is non-trivial, since the 
Josephson effect between two consecutive TS stripes becomes 
important. An additional term $\mathcal{F}_J$ 
should be added to the total free energy. 
\begin{equation}
\label{eqn:josephson-coupling}
\mathcal{F}_J = 
\sum_{n} \int_{overlap} dx 
J \vert D_{c, (n+1)}(x) - D_{c,n}(x) \vert^2,
\end{equation}
where the summation runs over all TS stripes. 
By adding this term,  
another line ($P_3$ in Fig. 1 inset) will emerge to 
denote a 2D $\to$ 3D superconductor crossover. 
Detailed calculations on refinement involving $\mathcal{F}_J$ will be 
the topic of a future publication.

{\it Summary:} We have proposed the possibility of coexistence
of antiferromagnetism and triplet superconductivity in 
the phase diagram of ${\rm (TMTSF)_2 PF_6}$. This intermediate phase
is proposed to be inhomogeneous and to consist of alternating 
insulating AF and TS stripes.
Two additional transition lines are present in a narrow 
range of pressures around $P_c$ separating the coexistence region
from the pure AF and pure TS phases. We estimate the maximum 
pressure range to be $\Delta P/P_c \approx 10\% $ at $T = 0$.

{\it Acknowledgements:}
We would like to thank NSF for support (DMR-0304380).

\end{document}